\documentstyle[prl,aps,epsfig]{revtex}         % 2 col mode
\begin{document}
\draft               % preprint mode
% 2 col mode:
\twocolumn[\hsize\textwidth\columnwidth\hsize\csname @twocolumnfalse\endcsname

\title{Semiconductor resonator solitons above band gap}
\author{V. B. Taranenko, C. O. Weiss, W. Stolz$^*$}
\address{Physikalisch-Technische Bundesanstalt 38116 Braunschweig, Germany\\
$^*$Philipps Universitaet, 35032 Marburg, Germany}
%\date{\today{}}
\maketitle
\begin{abstract}
We show experimentally the existence of bright and dark spatial
solitons in semiconductor resonators for excitation above the
band gap energy. These solitons can be switched on, both
spontaneously and with address pulses, without the thermal delay
found for solitons below the band gap which is unfavorable for
applications. The differences between soliton properties above
and below gap energy are discussed.
\end{abstract}
 \vskip1pc ]
% BEGIN TEXT HERE
We have recently shown that spatial solitons of various
characteristics can exist in nonlinear optical resonators
\cite{tag:1}, confirming various theoretical predictions
\cite{tag:2}. Our motivation for such investigations is the
common property of bistability and mobility of all such optical
spatial solitons, possibly useful in optical parallel information
processing. For such applications semiconductor resonators are
best suited. Technically the resonator structures needed are
rather precisely the structures used today for vertical-cavity
surface-emitting lasers (VCSEL), i.e. a stack of quantum wells
between Bragg mirrors.

For such structures in the dispersive limit we found the
hexagonal structure formation \cite{tag:3} theoretically
predicted for nonlinear dispersive resonators \cite{tag:4} and
demonstrated local switching of individual spots of the structure
with address pulses \cite{tag:3}. Closer to the band gap of the
quantum well material we were recently able to show the existence
of bright and dark spatial solitons \cite{tag:5}. The dynamics of
formation of these solitons \cite{tag:5} observed during
experiments for switching bright solitons on and off by local
addressing \cite{tag:6} was, however, found dominated by
dissipation and occurs thus with characteristic times of $\sim$ 1
$\mu$s, somewhat slow for most applications. We find here that
when working inside the absorption band, i.e. above the band gap
energy, (i) bright and dark solitons do also exist and (ii) can
be switched on apparently without the delay due to thermal
dissipation.\\

For the experiments on structures and solitons below or near band
gap we had used a resonator structure originally grown for
experiments on dispersive optical bistability. The resonance
wavelength of the resonator varies over the sample area from 30
nm longer than, to close to the band gap wavelength.

For the experiments inside the absorption band we use a structure
originally grown for use as an optically pumped picosecond VCSEL
\cite{tag:7}. For this sample the resonator resonance lies 10 nm
shorter than the band gap (exciton-) wavelength. The reflectivity
of the Bragg mirrors is 99,5 $\%$ and consequently, due to the
high linear absorption of the QW-material the resonance can not
be detected linearly. We found the resonance wavelength for a
particular point of the sample by applying a high laser field and
observing the (nonlinear) response.

For the measurements a set-up was used as described in
\cite{tag:5,tag:6}. It consists briefly of a \textit{cw} tunable
Ti:Al$_2$0$_3$-laser of a few hundred mW power, permitting to
illuminate an area of 50 to 100 $\mu$m diameter on the sample, and
detection equipment for time-resolved recording of the light
reflected from the sample. This is a camera allowing to take
snapshot pictures with a time resolution of 10 ns, and a small
area photodiode with 1 ns response, which could be imaged into
particular points of the illuminated area, thus allowing to
record the reflected intensity as a function of time, locally.
Mostly this photodiode is used to construct "streak"-images of
the intensity on a diameter across the illuminated area as
described in \cite{tag:3}, to visualize the soliton formation
dynamics. A tightly focused "address"-beam (of spot diameter 8
$\mu$m) split off from the illumination beam is used to apply
locally short switching pulses (20 ns). As the polarisation of the
address beam is perpendicular to that of the illumination beam,
the switching is incoherent (i.e. by local carrier generation,
and not by interference of light). Although the sample used was
designed as a laser, it showed bistability readily, as can be
expected for the parameters of the structure \cite{tag:7}.

\begin{figure}[htbf] \epsfxsize=82mm
\centerline{\epsfbox{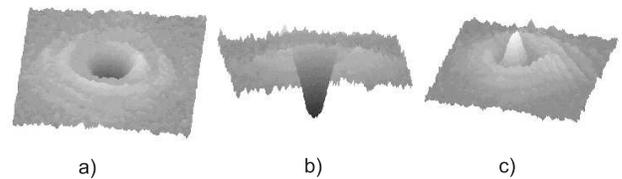}} \vspace{0.7cm} \caption{3D
representation of reflectivity. Bright soliton above band gap (a,
b). Due to observation in reflection the bright soliton appears
dark. a): view from above, b): from below, c): dark soliton.}
\end{figure}

Analogously to the excitation below/near band gap \cite{tag:5} at
smaller illumination bright solitons form and at higher intensity
dark solitons. Fig.~1 (a,b) shows the bright solitons as observed
in the reflected light with their characteristic concentric rings
\cite{tag:8} with the same appearance as the bright solitons
below band gap. Dark solitons (Fig.~1 c) also look similar to the
excitation below band gap \cite{tag:5}. As to be expected the
laser field had to be blue-detuned with respect to the resonator
to permit structure formation by "tilted waves" \cite{tag:9}.

\begin{figure}[htbf]
\epsfxsize=87mm \centerline{\epsfbox{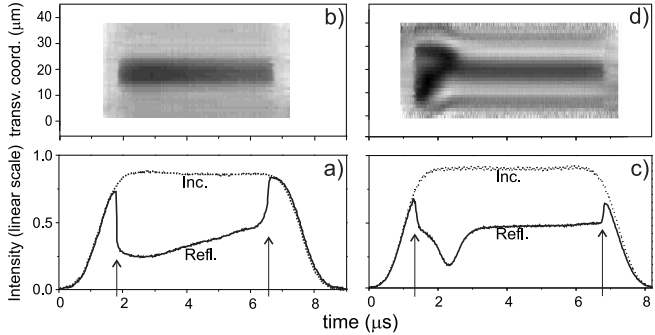}} \vspace{0.7cm}
\caption{Comparison of bright soliton formation above band gap
(left) and below band gap (right). b), d) reflectivity on a
diameter of the illuminated area as a function of time. a) c)
intensity of incident (dotted) and reflected (solid) light, at
the center of the soliton as a function of time. Arrows mark the
switch-on and -off.}
\end{figure}

Fig.~2 compares the dynamics of the bright soliton formation for
excitation above band gap (left) and below band gap (right). For
the latter case the solitons form (spontaneously) in the
following way: When the illumination intensity is increased, a
larger area of the resonator is switched rapidly. This switched
area contracts subsequently slowly to form the stable soliton.
When the intensity is reduced, the soliton is switched-off, as
one would expect.

This relatively slow soliton formation can be understood using a
simple model (similar to \cite{tag:10}) for the intracavity
optical field $E$ and carrier density $N$:
\begin{equation}
\cases{{\partial E}/{\partial t}=E_{\rm in}-E[1+2C{\rm
Im}(\alpha)(1-N)]-\cr
\quad\quad\quad\quad\quad\quad\quad\quad-iE[\theta-2C{\rm
Re}(\alpha)N-\nabla^{2}_{\bot}]\,,\cr \cr{\partial N}/{\partial
t}=-\gamma[N-|E|^2(1-N)-d\nabla^{2}_{\bot}N]\,,}
\end{equation}
where Re($\alpha$)$N$ is used to describe the reactive
nonlinearity. All other notations as in \cite{tag:10}. From (1)
we obtain the reflected intensity as a function of incident
intensity for wavelengths above (Re($\alpha$)$>$0), as well as
below the band gap (Re($\alpha$)$<$0), for plane waves (Fig.~3).
One sees that the bistability range is large below and small
above the band gap. Solving (1) numerically the typical bright
soliton (top of Fig.~3) is found coexisting with homogeneous
intensity solutions in the shaded regions of Fig.~3 a,b.

After switching on the resonator below band gap (Fig.~3b), the
intensity in the resonator is high and with it the thermal
dissipation. The temperature consequently rises, which shifts the
band gap \cite{tag:11} and with it the bistability
characteristic, so that the switch-off intensity, close to which
the stable solitons exist, becomes close to the incident
intensity.  Then the resonator is in the basin of attraction for
the solitons and the soliton forms as observed in Fig.~2 c,d.

\begin{figure}[htbf]
\epsfxsize=87mm \centerline{\epsfbox{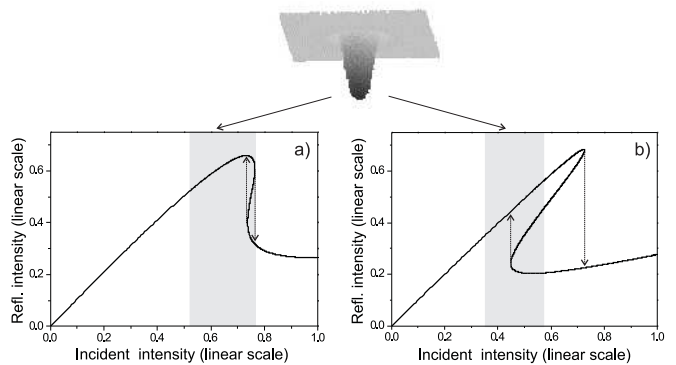}} \vspace{0.7cm}
\caption{Steady-state plane wave solution of Eq.(1) above band
gap a): $\rm Re(\alpha)=0.05$; and below band gap b): $\rm
Re(\alpha)=-0.05$. Other parameters: $C=15$, $\rm
Im(\alpha)=0.99$, $\theta=-3$, $d = 0.1$. The soliton solution
shown exists for incident intensities corresponding to the shaded
areas, in coexistence with homogeneous solutions. For a
temperature increase the characteristics together with soliton
existence ranges shift to higher incident intensities. Reflected
and incident intensities normalized to the same value. }
\end{figure}

\begin{figure}[htbf]
\epsfxsize=81mm \centerline{\epsfbox{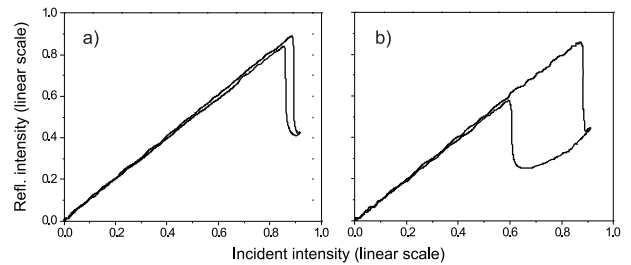}} \vspace{0.7cm}
\caption{Bistability characteristics (reflected light intensity
versus incident light intensity) measured at the center of an
illuminated area of 100 $\mu$m diameter, for above band gap (a))
and below band gap (b)) excitation. }
\end{figure}

Above the band gap Fig.~2 a,b  show that the soliton is switched
on immediately without the slow thermal process. Fig.~3a shows
why. The plane wave characteristic of the resonator above band
gap is either bistable but very narrow, or even monostable (due
to the contribution of the self-focusing reactive nonlinearity
\cite{tag:12}) but still with bistability between the soliton
state (not plane wave) and the unswitched state. In this case the
electronic switching leads directly into the basis of attraction
for solitons and the switch-on of the soliton is purely electronic
and fast. The width of the bistability characteristics observed
experimentally (Fig.~4) scale in agreement with Fig.~3.

Nonetheless, also above band gap there is strong dissipation after
the switch-on. The associated temperature rise influences and can
even destabilize the soliton. The effect can be seen in Fig.~2a.
Over a time of a few $\mu$s after the soliton switch-on the
soliton weakens (reflectivity increases slowly) presumably by the
rise of temperature and the associated shift of the band gap. At
6.5 $\mu$s the soliton switches off although the illumination has
not yet dropped.

\begin{figure}[htbf]
\epsfxsize=60mm \centerline{\epsfbox{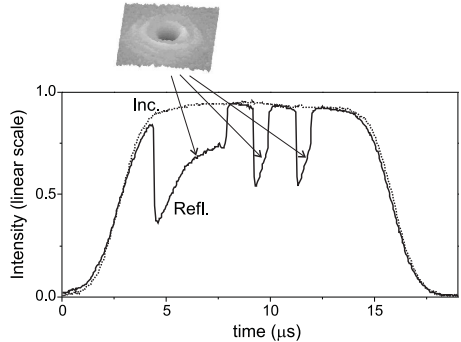}} \vspace{0.7cm}
\caption{Spontaneous repeated switching on and off of a bright
soliton due to thermal dissipation. }
\end{figure}

\begin{figure}[htbf]
\epsfxsize=65mm \centerline{\epsfbox{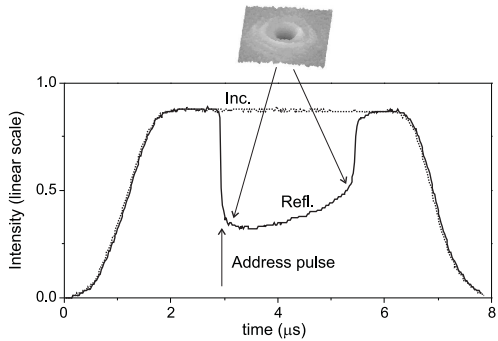}} \vspace{0.7cm}
\caption{Switching on an above band gap soliton with an address
pulse. Arrow marks application of the pulse. Incident light at
soliton center, dotted; reflected light intensity at soliton
center, solid. }
\end{figure}

Thus, while the dissipation does not hinder the fast switch-on of
the soliton, it finally destabilizes the soliton. After the
soliton is switched off, the material cools and the band gap
shifts back so that the soliton could switch on again. Fig.~5
confirms this by observing for a longer time.

Fig.~6 shows that the switch-on of a soliton is fast not only
when the soliton appears spontaneously, but also when a soliton
is switched on by an address beam. The maximum intensity of the
illumination is kept here below the value for spontaneous
appearance of solitons (but above the intensity below which a
soliton disappears). At t = 3 $\mu$s the address pulse (50 ns
duration, polarized perpendicularly to the illumination, 8 $\mu$m
diameter) is applied. As can be seen the switch-on of a soliton
is fast. Again, as in Figs~2a and 5, heating appears to
destabilize subsequently the soliton.

It might appear at this point that the price to pay for a fast
electronic switch-on of solitons is their longer term
destabilization. However, this must not be the case. In the
experiments with excitation below the band gap the heating plays
actually a constructive role in the soliton formation.
Consequently one could hope that parameters can be found where
the bistability characteric of the resonator is narrow, but the
temperature shift of the band gap would aid or at least not
affect the soliton stability. Such is in fact possible as shown
in Fig.~7. The switch-on of the soliton is direct and the soliton
does not destabilize over time.\\

\begin{figure}[htbf]
\epsfxsize=65mm \centerline{\epsfbox{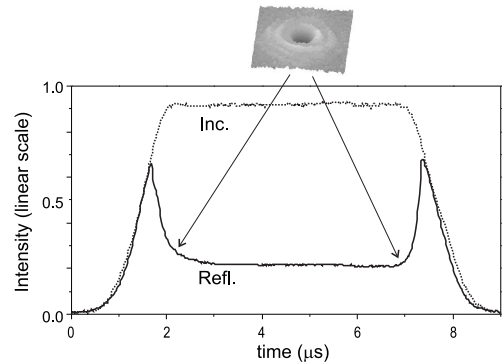}} \vspace{0.7cm}
\caption{Spontaneous direct switch-on of above band gap soliton
which remains stable in the presence of heating. The initial
switch-on of soliton ($t$ = 1.8 $\mu$s) is fast. The further
decrease of reflected intensity (i.e. increase of intracavity
soliton field) follows adiabatically the incident intensity. }
\end{figure}

In summary, we find that for excitation above the semiconductor
material band gap spatial resonator solitons exist and can be
directly switched on without the delay due to thermal
dissipation, as it is found for excitation below the band gap.
Conditions can be found where this direct switch on is not
accompanied by destabilization of the solitons due to the
heating. Thus, fast switching of spatial solitons as required for
applications is compatible with stable solitons even with strong
heat dissipation. We understand the different soliton formation
dynamics above and below band gap in terms of the different
widths of the resonator plane wave bistability characteristics.
The structure used for the observation of above band gap solitons
was a vertical cavity laser.\\

Acknowledgement\\ This work was supported by ESPRIT LTR project
PIANOS. We gratefully acknowledge discussions with K.Staliunas.

\end{document}